\def\drft{\true}
\ifx \drft \undefined
\documentclass[onecolumn,12pt]{jetpl}
\else
\documentclass{jetpl}
\twocolumn
\fi
\usepackage{epstopdf}
\usepackage[]{hyperref}

\def\d{\partial}

\newcommand{\be}{\begin{equation}}
\newcommand{\ee}{\end{equation}}
\newcommand{\bea}{\begin{eqnarray}}
\newcommand{\eea}{\end{eqnarray}}
\newcommand{\bg}{\begin{gather}}
\newcommand{\eg}{\end{gather}}
\newcommand{\bseq}{\begin{subequations}}
\newcommand{\eseq}{\end{subequations}}

\lat

\title{Towards conformal cosmology}

\rtitle{Towards conformal cosmology}

\sodtitle{Towards conformal cosmology}

\author{
M.\,Libanov$^{+}$,
\/\thanks[1]{e-mail: ml@ms2.inr.ac.ru}
V.\,Rubakov$^{+*}$,
\/\thanks[2]{e-mail: rubakov@ms2.inr.ac.ru}
G.\,Rubtsov$^{+}$
\/\thanks[3]{e-mail: grisha@ms2.inr.ac.ru}}

\rauthor{M.\,Libanov, V.\,Rubakov, G.\,Rubtsov}

\sodauthor{Libanov, Rubakov, Rubtsov}

\address{
$^+$
Institute for Nuclear Research of
         the Russian Academy of Sciences,\\  60th October Anniversary
  Prospect, 7a, 117312 Moscow, Russia
\\~\\
$^*$
Department of Particle Physics and Cosmology,
Physics Faculty, Moscow State University\\ Vorobjevy Gory,
119991, Moscow, Russia
}

\abstract{Approximate
de Sitter symmetry of inflating Universe is responsible for
 the approximate flatness of 
the power spectrum of scalar perturbations. However, this
is not the only option. Another symmetry which can explain nearly
scale-invariant power spectrum is conformal invariance. We give a
short review of models based on conformal symmetry which lead to
the scale-invariant spectrum of the scalar perturbations. We discuss also
potentially observable features of these models.}

\begin{document}
\maketitle

Observational data show that primordial scalar
perturbations in the Universe must have been generated 
at some early cosmological
stage, preceding the hot epoch. They are nearly Gaussian and have
nearly flat power spectrum~\cite{Hinshaw:2012aka}. 
The first property suggests that these perturbations originate
from amplified vacuum fluctuations of weakly coupled quantum field(s).
Indeed, the defining property of Gaussian random field $\zeta
(\mathbf{x})$ is that it obeys the  Isserlis--Wick 
theorem, which holds also for
any free quantum field in its
vacuum state, while linear evolution in classical
background does not induce non-Gaussianity.

The second property is also very suggestive.
The power spectrum $\mathcal{ P}(k)$ defined as
\[
\langle \zeta (\mathbf{k})\zeta (\mathbf{k}'\rangle = \frac{1}{4\pi
k^{3}}\mathcal{ P}(k)\delta (\mathbf{k}+\mathbf{k}')
\]
gives the fluctuation in logarithmic interval of momenta,
\[
\langle (\zeta (\mathbf{x}))^{2}\rangle =\int \limits_{0}^{\infty
}\frac{dk}{k}\mathcal{ P}(k),\ \ \ \mathcal{ P}\propto k^{n_{s}-1},
\]
where $n_{s}$ is a spectral index. The flat or scale invariant spectrum
corresponds to $n_{s}=1$. In early 70's Harrison, 
Zeldovich and Peebles and Yu~\cite{Harrison:1969fb} 
conjectured that the spectrum is flat, to avoid
large deviation from homogeneity
and isotropy of the observed Universe on large scales and black hole
formation on small scales. Current observational
data~\cite{Hinshaw:2012aka}  show that the power spectrum is indded
nearly flat
and give $n_{s}-1\approx -0.032$.

The flatness of the power spectrum may be due to some symmetry. The best
known candidate is the symmetry $SO(4,1)$ of the de Sitter metric
$ds^{2}=dt^{2}-\mathrm{e}^{2Ht}d\mathbf{x}^{2}$, which includes
 spatial dilatations
supplemented by time translations: $\mathbf{x}\to \lambda \mathbf{x}$,
$t\to t-(H)^{-1}\log\lambda$. This is the approximate symmetry of 
inflating Universe~\cite{inflation}, and, indeed, the inflationary
mechanism of the generation of scalar
perturbations~\cite{infl-perturbations} produces almost flat power
spectrum. Despite this success of the inflationary paradigm,
search for alternatives to inflation in general and to de~Sitter
symmetry in particular is of obvious interest. Alternatives to inflation
include ekpyrotic model~\cite{ekpyrosis} with negative exponential potential
and bounce~\cite{minus-exp}, 
"starting"~\cite{starting} Universe, etc,
while there are several
mechanisms capable of producing flat or almost flat scalar
spectrum~\cite{Wands:1998yp,Allen,Mukohyama:2009gg,minus-old}. 
In some cases,
there is no obvious symmetry that guarantees the flatness, i.e., the
scalar spectrum is flat accidentally.

In quest for an alternative symmetry behind the nearly
flat scalar spectrum one
naturally turns to conformal symmetry $SO(4,2)$~\cite{Rubakov:2009np,
Creminelli:2010ba, Hinterbichler:2011qk, Hinterbichler:2012fr}  (see
Ref.~\cite{Antoniadis:1996dj} for related discussion). Conformal group
includes dilatations, $x^{\mu }\to \lambda x^{\mu }$, 
which in the end may be responsible for the scale-invariant
scalar spectrum. An assumption of conformal invariance at the time the
primordial perturbations are generated is in line with the viewpoint that
the underlying theory of Nature may have conformal phase, and that the
Universe may have started off from, or passed through an unstable
conformal state and then evolved to much less symmetric state we see today.

In this paper we give a short review of models based on  conformal
symmetry, which  lead to the scale-invariant spectrum of the scalar
perturbations. It is worth noting that one does not necessarily have 
to consider
these models as alternatives to inflation, since some of them can work in
inflating Universe as well.

\section{Two sample ways of getting flat scalar spectrum}
\label{Section/Pg2/1:cc_jetpl/Two ways}
\subsection{Conformal and global symmetry instead of de Sitter symmetry}
\label{confroll}

To begin with, let us consider a scenario proposed in
Ref.~\cite{Rubakov:2009np}. In this scenario conformal symmetry is
supplemented by a global symmetry. The simplest model of this sort has
global symmetry $U(1)$ and involves complex scalar field $\phi $, which is
conformally coupled to gravity. The action of the model is
$S=S_{G+M}+S_{\phi }$, where $S_{G+M}$ is the action of gravity and
dominating matter, while the non-trivial dynamics of the scalar field, which is
assumed to be spectator, is
governed by 
\[
S_{\phi } = \int d^4x \sqrt{-g} \left[ g^{\mu\nu}\partial_\mu\phi^*
\partial_\nu\phi + \frac{R}{6} \phi^* \phi - (-h^{2}|\phi |^{4}) \right] \; .
\]
Thus,  quartic potential 
allowed by conformal
invariance is assumed to be \textit{negative}.
Therefore, $\phi =0$ is an unstable state with unbroken conformal
symmetry. One assumes that the background space-time is homogeneous,
isotropic and spatially flat,  $ds^{2}=a^{2}(\eta )(d\eta
^{2}-d\mathbf{x}^{2})$. Then in terms of the field $\chi (\eta
,\mathbf{x})=a(\eta )\phi (\eta ,\mathbf{x})$ the dynamics is the same as
in flat space-time,
\begin{equation}
\eta ^{\mu \nu }\partial _{\mu }\partial _{\nu }\chi -2h^{2}|\chi
|^{2}\chi =0.
\label{Eq/Pg3/1A:cc_jetpl}
\end{equation}
Spatially homogeneous background approaches the late-time attractor
\begin{equation}
\chi _{c}(\eta )=\frac{1}{h(\eta _{*}-\eta )},
\label{Eq/Pg3/1:cc_jetpl}
\end{equation}
where $\eta _{*}$ is an arbitrary real parameter, and
we consider real solution, without loss of generality. The background
solution (\ref{Eq/Pg3/1:cc_jetpl}) breaks conformal group $SO(4,2)\to
SO(4,1)$.  The meaning of the parameter $\eta _{*}$
is that the field $\chi _{c}$ would run away to infinity as $\eta \to \eta
_{*}$, if the scalar potential remained negative quartic at arbitrarily
large fields. It is worth noting that the particular behaviour $\chi
_{c}\propto (\eta _{*}-\eta )^{-1}$ is dictated by conformal symmetry.

\paragraph{Phase perturbations.} To see how the scale invariant spectrum
emerges in the model,  let us consider perturbations of the phase $\theta
=\mathrm{Arg}\phi $, or, for the real background
(\ref{Eq/Pg3/1:cc_jetpl}), perturbations of the imaginary part $\chi
_{2}\equiv \mathrm{Im}\chi /\sqrt{2}$. At the
linearized level, perturbations of the phase and modulus decouple 
and the linearized equation is
\begin{equation}
(\delta \chi_2)^{\prime \prime}
- \partial_i \partial_i \; \delta \chi_2
- \frac{2}{(\eta_* - \eta)^2} \; \delta \chi_2  = 0 \; .
\label{Eq/Pg3/2:cc_jetpl}
\end{equation}
An important assumption of the entire scenario is that the rolling stage
begins early enough, so that there is time at which the following
inequality holds:
\begin{equation}
k(\eta _{*}-\eta )\gg 1,
\label{Eq/Pg3/3:cc_jetpl}
\end{equation}
where $k=|\mathbf{k}|$ is conformal momentum. Since the momenta $k$ of
cosmological significance are as small as the present Hubble parameter,
this inequality means that the duration of the rolling stage in conformal
time is longer than the conformal time elapsed from, say, the beginning of
the hot Big Bang expansion to the present epoch. This is only possible if
the hot Big Bang stage was preceded by some other epoch, at which the
standard horizon problem is solved; the mechanism we discuss here is meant
to operate at that epoch. We note in passing that the latter property is
inherent in most, if not all, mechanisms of the generation of cosmological
perturbations.

Equation~(\ref{Eq/Pg3/2:cc_jetpl}) is exactly the same as equation for
minimally coupled massless scalar field in the
de~Sitter background. Nevertheless, let us briefly discuss its solutions.
At early times, when the inequality (\ref{Eq/Pg3/3:cc_jetpl}) is
satisfied, the third term in (\ref{Eq/Pg3/2:cc_jetpl}) is negligible and
$\delta \chi_2$ is free massless quantum field,
\[
\delta \chi_2({\bf x}, \eta) = \int~\frac{d^3k}{(2\pi)^{3/2}
\sqrt{2k}}~\left( \delta \chi_2^{(-)}({\bf k},
{\bf x}, \eta)  \hat{A}_{\bf k} + h.c.\right)\; ,
\]
whose modes are
\begin{equation}
\delta \chi_2^{(-)} ({\bf k},
{\bf x}, \eta)=
\mbox{e}^{i {\bf k x} - ik\eta}
\; .
\label{Eq/Pg4/1:cc_jetpl}
\end{equation}
Here  $\hat{A}_{\bf k}$ and  $\hat{A}_{\bf k}^\dagger$  are annihilation
and creation operators obeying the standard commutational relation,
$[\hat{A}_{\bf k},\hat{A}_{\bf k^\prime}^\dagger] = \delta({\bf k} - {\bf
k^\prime})$. It is  natural to assume that the field $\delta \chi_2$ is
initially in its vacuum state.

The rolling background $\chi_c (\eta)$ produces an effective ``horizon''
for the perturbations $\delta \chi_2$. The oscillations
(\ref{Eq/Pg4/1:cc_jetpl}) terminate when the mode exits the ``horizon'',
i.e., at $k(\eta_* - \eta) \sim 1$. The solution to
eq.~(\ref{Eq/Pg3/2:cc_jetpl})
with the initial condition
(\ref{Eq/Pg4/1:cc_jetpl}) is
\begin{equation}
\delta \chi_2^{(-)} ({\bf k},
{\bf x}, \eta)= \mbox{e}^{i {\bf kx} - ik\eta_*}\cdot F(k, \eta_* - \eta)
\; ,
\label{Eq/Pg4/2:cc_jetpl}
\end{equation}
where
\begin{equation}
F(k, \xi) = - \sqrt{\frac{\pi}{2}k \xi}~
H^{(1)}_{3/2} (k\xi)
\label{Eq/Pg4/3:cc_jetpl}
\end{equation}
and $H^{(1)}_{3/2}$ is the Hankel function. In the late-time
super-''horizon'' regime, when $k(\eta_* - \eta) \ll 1$ and the third
term in (\ref{Eq/Pg3/2:cc_jetpl}) dominates, one has
\begin{equation}
F(k, \eta_* - \eta) = \frac{i}{k(\eta_* - \eta)} \; .
\label{Eq/Pg5/1:cc_jetpl}
\end{equation}
Hence, the super-''horizon'' perturbations of the phase
$\delta \theta \equiv \delta \chi_2/\chi_c$ are time-independent,
\begin{equation}
\delta \theta ({\bf x}) =\frac{\delta \chi_2({\bf x}, \eta)}{\chi_c
(\eta)} = ih \int~\frac{d^3k}{4\pi^{3/2}k^{3/2}}~
 \mbox{e}^{i{\bf kx} - i k\eta_*}
\hat{A}_{\bf k}
+ h.c.
\label{Eq/Pg5/2:cc_jetpl}
\end{equation}
This expression describes Gaussian random field 
whose power spectrum is flat:
\begin{equation}
\mathcal{ P}_{\delta \theta } =\frac{h^2}{(2\pi)^2 }.
\label{Eq/Pg5/3:cc_jetpl}
\end{equation}

We emphasize that this result is an automatic consequence of the
global $U(1)$ and conformal symmetries. To see this, let us consider long
wavelength regime. In this regime the second term in
(\ref{Eq/Pg3/2:cc_jetpl}) is negligible and (\ref{Eq/Pg3/2:cc_jetpl})
becomes the equation for spatially homogeneous perturbation. Recall that
$\chi _{c}$ is the spatially homogeneous solution to the full field equation
(\ref{Eq/Pg3/1A:cc_jetpl}), hence, due to $U(1)$ symmetry,
$\mathrm{e}^{i\alpha }\chi _{c}$
is also a solution, where $\alpha $ is a real constant. For small $\alpha $
the latter solution is $\chi _{c}+i\alpha \chi _{c}$, and the imaginary
part is a small perturbation, which is precisely $\delta \chi _{2}$. So, if
the perturbation $\delta \chi _{2}$ oscillates with unit amplitude, it
behaves at late times as (cf. (\ref{Eq/Pg4/1:cc_jetpl}) and 
(\ref{Eq/Pg5/1:cc_jetpl})) $\delta \chi _{2}=C/[k(\eta _{*}-\eta )]$,
where the factor $k^{-1}$ is evident on dimensional grounds and $C$ is
independent of time and $k$.

Deviation from exact conformal invariance naturally gives rise to the tilt
in the power spectrum, which depends both on the way conformal invariance
is broken and on the evolution of the scale factor~\cite{Osipov:2010ee}.

\subsection{Galilean Genesis}

The Galileon model has been introduced in Ref.~\cite{Nicolis:2008in}.
In Minkowski space-time, the Lagrangian of the
simplest conformally-invariant version~\cite{Creminelli:2010ba} of the model
 is
\begin{equation}
L_\pi =  f^2 {\rm e}^{2\pi} \partial_\mu \pi \partial^\mu \pi + \frac{f^3}{\Lambda^3}
\partial_\mu \pi \partial^\mu \pi \Box \pi + \frac{f^3}{2\Lambda^3} (\partial_\mu \pi \partial^\mu
\pi)^2 \; ,
\label{Eq/Pg6/1:cc_jetpl}
\end{equation}
where $\Box = \partial_\mu \partial^\mu$. This Lagrangian is conformally 
invariant,
with $\pi $ transforming under dilatations as $\mathrm{e}^{\pi (x)}\to
\lambda \mathrm{e}^{\pi (\lambda x)}$.

In the Galilean Genesis scenario the Universe begins from Minkowski
space-time. The field equation in Minkowski space-time admits a
homogeneous attractor solution
\begin{equation}
\mathrm{e}^{\pi_c} =  \frac{1}{H_G (t_{*}-t)} \; ,
\label{Eq/Pg6/2:cc_jetpl}
\end{equation}
where $ H_G^2 = 2\Lambda^3/(3 f) $. The form of the solution is
again dictated by conformal invariance.

Initial energy density is zero, while effective pressure is negative.
Then the energy density slowly builds up and the Hubble
parameter grows in time,
\[
H(t)=\frac{1}{3}\frac{f^{2}}{M_{Pl}^{2}}\frac{1}{H_G^2 (t_* - t)^3} \; ,
\]
until $(t_{*}-t)\sim H_{G}^{-1}\cdot f/M_{Pl}$. The growth of the
Hubble parameter is due to violation of
all energy conditions.
Nevertheless, the theory is fully self-consistent: there are
no ghosts, tachyons, and
other pathologies (there is, however, superluminality issue which has 
not been quite settled~\cite{superlum}). At some time Galileon is assumed to transmit its energy
to conventional matter, and hot epoch begins.

Galileon perturbations per se are not  suitable for generating scalar
perturbations (see discussion in Sec.~\ref{modulusperturbations}).  For the
purpose of generating the scalar perturbations another field $\theta $ of
conformal weight 0 is introduced.  By conformal invariance, its quadratic
Lagrangian has the form
\[
L_{\theta }=\mathrm{e}^{2\pi }(\partial _{\mu }\theta )^{2} \ \
\Rightarrow \ \ L_{\theta }(\pi
_{c})=\frac{\mathrm{const}}{(t_{*}-t)^{2}}\cdot(\partial _{\mu }\theta
)^{2}.
\]
That is, the dynamics of perturbations $\delta \theta $ in the background
$\pi _{c}$ is exactly the same as in the conformal rolling model discussed
above.

The similarity between Galilean Genesis and conformal rolling model is not
accidental. In Ref.~\cite{Hinterbichler:2011qqk}
general arguments are given, which show that the scale invariant power
spectrum is inherent in an entire class of models. The general setting is
conformally invariant theory of a scalar field $\rho $ of
conformal weight $\Delta\neq 0$ in effectively Minkowski
space-time. Up to rescaling this field corresponds to $|\phi |$ in the
conformal rolling model and to $\mathrm{e}^{\pi }$ in the Galilean Genesis
scenario; in both models $\Delta=1$. The form of homogeneous
classical solution
\begin{equation}
\rho _{c}(t) =\frac{1}{(t_{*}-t)^{\Delta}}
\label{Eq/Pg7/1:cc_jetpl}
\end{equation}
is dictated by conformal invariance ($t$ is conformal time in the case
of the conformal rolling scenario). As mentioned above,
 the perturbations of the field
$\rho $ do not have flat power spectrum, so another
\textit{spectator} scalar field $\theta $ of conformal weight 0 is
introduced. Then by conformal invariance,
the kinetic term in the Lagrangian of $\theta $ is
 $ L_{\theta }\propto \rho
^{2/\Delta}(\partial _{\mu }\theta )^{2}$. Assuming that  possible
potential terms are negligible the Lagrangian in the rolling background
(\ref{Eq/Pg7/1:cc_jetpl}) takes the form
\[
L_{\theta }=\frac{\mbox{const}}{(t_{*}-t)^{2}} (\partial _{\mu }\theta )^{2},
\]
which is exactly the same as the Lagrangian of
a scalar field minimally coupled to gravity in 
de~Sitter space with conformal time $t$ and scale factor $a(t)\propto
1/(t_{*}-t)$. As a result, $\theta $ develops perturbations with the flat
power spectrum. 

\subsection{Further aspects}
\label{Reprocess}

Obviously, generating the field perturbations $\delta \theta$
is not the whole story. There are several other ingredients
of the conformal scenario. Some of them have not yet been worked out
in detail.

\paragraph{Beginning of rolling.} The rolling stage 
(\ref{Eq/Pg7/1:cc_jetpl}) has to start in one or another way.
One possibility is spontaneous decay of the unstable
conformally invariant vacuum $\rho = 0$, which proceeds through bubble
nucleation. Such a decay has been discussed in Ref.~\cite{Libanov:2015mha}
within the holographic approach (holographic picture of
the conformal scenario has been suggested earlier in 
Refs.~\cite{Hinterbichler:2014tka,mlvr-holo}). Even though the rolling field is
not spatially homogeneous in the false vacuum decay process, 
the field perturbations
have the properties we discuss in Secs.~\ref{confroll}, 
\ref{modulusperturbations}.

\paragraph{End of rolling.}  The rolling stage 
(\ref{Eq/Pg7/1:cc_jetpl}) should terminate at some late time.
This implies that conformal invariance is broken at large
field values. As an example, in the conformal rolling scenario of
Sec.~\ref{confroll} one assumes that the scalar potential has a minimum
or nearly flat valley at large $|\phi|$ and that  $|\phi|$
eventually settles there. Furthermore, in the Galilean Genesis
scenario there must be a stage of defrosting, i.e., transmission
of energy from the rolling field $\rho$ to 
heat~\cite{LevasseurPerreault:2011mw}, after which the usual hot Big Bang
epoch begins.

\begin{figure}[tb!]
\begin{center}
\includegraphics[width=0.4\textwidth,angle=0]{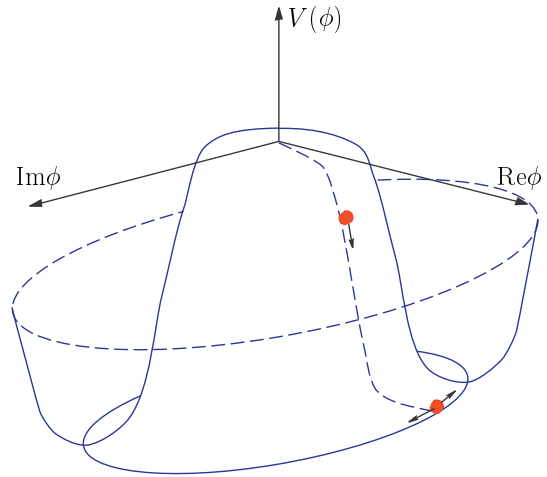}
\end{center}
\caption{{\rm Fig. 1}. The scalar potential in the pseudo-Nambu--Goldstone
scenario.
\label{fig1}
}
\end{figure}

\paragraph{Reprocessing perturbations $\delta \theta$
into adiabatic perturbations.} The field perturbations
 $\delta \theta$ are to be converted into adiabatic perturbations.
This can happen at the hot Big Bang epoch. One possibility
is to make use of the curvaton mechanism~\cite{Linde:1996gt}.
As an example, in the scenario of Sec.~\ref{confroll} the phase
$\theta$
may actually be a pseudo-Nambu--Goldstone
field. Generically, conformal rolling ends up at a slope of its potential,
see Fig.~\ref{fig1}.
The perturbations $\delta \theta$ are reprocessed into adiabatic
perturbations at later time at the hot Big Bang epoch when
the field $\theta$ oscillates and decays into conventional particles,
cf. Ref.~\cite{Dimopoulos:2003az}.

Another possibility is the modulated decay mechanism suggested
in the inflationary context in 
Refs.~\cite{Dvali:2003em,Dvali:2003ar,Vernizzi:2003vs}.

In both cases the conversion of the field perturbations $\delta \theta$
into adiabatic perturbations induces some degree of non-Gaussianity.
This is not a specific property of the conformal scenario, however,
as it holds in all models employing the curvaton or modulated
decay mechanism, including versions of  inflation.

\section{Perturbations of modulus}
\label{modulusperturbations}

Let us now come back to the conformal rolling stage. From now on 
we use the nomenclature of the 
negative quartic model of Sec.~\ref{confroll} for definiteness.
We are interested in the
radial perturbations~\cite{Rubakov:2009np, Libanov:2010nk, Libanov:2010ci},
i.e., perturbations of the modulus of the field $\chi$, or, with our
convention of real background $\chi_c$, perturbations of the real part $
\chi_1 \equiv \mbox{Re}~\chi/{\sqrt{2}} $. At the linearized level, they
obey the following equation,
\[
(\delta \chi_1)^{\prime \prime} - \partial_i\partial_i~ \delta \chi_1 -
6 h^2 \chi_c^2 \delta \chi_1
= 0 \; .
\]
Its solution that tends to properly normalized mode of free quantum
field at early times, $k(\eta_* -\eta) \to \infty$, is
\[
\delta \chi_1 = - \mbox{e}^{i{\bf kx}-ik\eta_*} \cdot \frac{i}{4\pi}
\sqrt{\frac{\eta_* - \eta}{2}} H_{5/2}^{(1)} \left[k (\eta_* - \eta)
\right] \cdot \hat{B}_{\bf k} + h.c.\; ,
\]
where $\hat{B}_{\bf k}$, $\hat{B}_{\bf k}^\dagger$ is another set of
annihilation and creation operators. At late times, when $k(\eta_* - \eta)
\ll 1$ (super-''horizon'' regime), one has
\[
\delta \chi_1 = - \mbox{e}^{i{\bf kx}-ik\eta_*} \cdot
\frac{3}{4\pi^{3/2}} \frac{1}{k^{5/2} (\eta_* - \eta)^2} \cdot
\hat{B}_{\bf k} + h.c.\; .
\]
Hence, the super-''horizon'' perturbations of the modulus have red
power spectrum
\begin{equation}
\mathcal{ P}_{|\phi |}(k)\propto k^{-2}.
\label{Eq/Pg10/1:cc_jetpl}
\end{equation}

The dependence $\delta \chi_1 \propto (\eta_* - \eta)^{-2}$ is 
interpreted in terms of the local shift of the ``end time'' parameter
$\eta_*$. Indeed, with the background field given by
(\ref{Eq/Pg3/1:cc_jetpl}), the sum $\chi_c + \delta \chi_1$,
i.e., the radial field including perturbations, is the linearized form of
\begin{equation}
\chi_c [\eta_* ({\bf x}) - \eta] = \frac{1}{h[\eta_* ({\bf x}) - \eta]} \;
,
\label{Eq/Pg10/2:cc_jetpl}
\end{equation}
where
$\eta_* ({\bf x}) = \eta_* + \delta \eta_* ({\bf x})$
and
\begin{equation}
\delta \eta_* ({\bf x}) =- \frac{3h}{4\pi^{3/2}}\int~\frac{d^3k}{k^{5/2}}
\left( \mbox{e}^{i{\bf kx} - ik\eta_*} \cdot \hat{B}_{\bf k} + h.c.
\right) \; .
\label{Eq/Pg10/4:cc_jetpl}
\end{equation}
So, the infrared radial modes modify the effective background by
transforming the ``end time'' parameter $\eta_*$  into 
time-independent random field that
slowly varies in space,
with red power spectrum clearly seen from
(\ref{Eq/Pg10/4:cc_jetpl}). This observation is valid beyond the
linear approximation: once the spatial scale of variation of $\chi_1 ({\bf
x},\eta)$ exceeds the ``horizon'' size, spatial gradients in
eq.~(\ref{Eq/Pg3/1A:cc_jetpl}) are negligible, and the late-time solutions
to the full non-linear field equation have locally one and the same form
(\ref{Eq/Pg3/1:cc_jetpl}), modulo slow variation of $\eta_*$ in space.

A few remarks are in order.
First, the infrared modes contribute both to the
field $\delta \eta_* ({\bf x})$ itself and to its spatial derivatives. The
contribution of the modes which are superhorizon today, i.e., have momenta
$k\lesssim H_0$, to the fluctuation of $\partial_i \eta_*$ is given by
\begin{eqnarray}
\langle \partial_i \eta_* ({\bf x}) \partial_j \eta_* ({\bf x})
\rangle_{k\lesssim H_0} &=& \delta_{ij} \cdot \frac{3h^2}{4\pi}
\int_{k\lesssim H_0} \frac{dk}{k} \nonumber\\
&=& \delta_{ij} \cdot \frac{3h^2}{4\pi} \log \frac{H_0}{\Lambda} \; ,
\label{Eqn/Pg11/1:cc_jetpl}
\end{eqnarray}
where $\Lambda$ is the infrared cutoff which parametrizes our ignorance of
the dynamics at the beginning of the conformal rolling stage.

Second, modulo field redefinition and notations, the
properties of Galileon perturbations are exactly the same as the
properties of
radial perturbations $\delta \chi _{1}$ in conformal rolling
scenario~\cite{Libanov:2011bk}. Furthermore, these properties are
unambiguously determined by  conformal invariance~\cite{Libanov:2011bk,
Hinterbichler:2011qqk}. The same properties -- flat and red spectra
of zero conformal weight and rolling fields, respectively -- are inherent
in the false vacuum decay setup mentioned in Sec.~\ref{Reprocess}.
Thus, we are dealing with the whole class of models.

Finally, so far we have considered  spectator  fields.
That is, we have ignored the backreaction of the scalar fields on gravity.
In particular, we have considered a scalar field \textit{conformally}
coupled to gravity (in the conformal rolling scenario of Sec.~\ref{confroll}) 
and assumed
negligible energy density (in both Galilean Genesis and conformal
rolling cases). However, it is of interest
to consider also  dynamical versions
with scalar fields \textit{minimally} coupled to gravity and dominating
the cosmological evolution. In that case there is a potential danger that the
strong-coupling regime arises at too low energy
scales~\cite{Hinterbichler:2011qk}. This option has
been studied in Ref.~\cite{Libanov:2011zy}. It has been shown that mixing of
the
scalar field(s) with the metric in dynamical pseudo-conformal models
does not introduce new strong-coupling UV scales. Furthermore,  the spectator
approximation gives correct results in dynamical models provided 
 that the background space-time is sufficiently flat . 
This applies, in particular, to  potentially observable effects discussed in
Sec.~\ref{Section/Pg11/1:cc_jetpl/Effect}.
These effects are inherent in the entire class of both
spectator and dynamical (pseudo-)conformal models.

\section{Effect of infrared radial modes on perturbations of phase}
 \label{Section/Pg11/1:cc_jetpl/Effect}

\begin{figure}[tb!]
\begin{center}
\includegraphics[width=0.4\textwidth,angle=0]{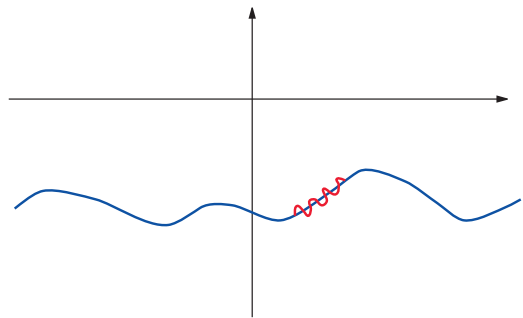}
\end{center}
\caption{{\rm Fig. 2}. Due to the perturbations of the radial field, the
evolution of phase perturbations proceeds in 
inhomogeneous background that slowly varies in space.
\label{fig2}
}
\end{figure}

Let us discuss how the interaction with the infrared radial
modes affects the properties of the phase perturbations  $\delta
\theta $~\cite{Libanov:2010nk, Libanov:2010ci, Libanov:2011hh}. To this
end, we consider perturbations of the imaginary part $\delta \chi _{2}$,
whose wavelengths are much smaller than the scale of the spatial variation
of the modulus. Because of the separation of scales, perturbations $\delta
\chi _{2}$ can still be treated in the linear approximation, but now in
the background (\ref{Eq/Pg10/2:cc_jetpl}), see Fig.~\ref{fig2}.

Since our concern is the infrared part of $\eta _{*}(\mathbf{x})$, we make
use of the spatial gradient expansion, consider a region near the origin
and write
\begin{equation}
\eta_* ({\bf x}) = \eta_* (0) - v_i x_i + \dots \; , \;\;\;\;\;
v_i = -\d_i \eta_*({\bf x})\vert_{{\bf x}=0} \; ,
\label{Eq/Pg12/1:cc_jetpl}
\end{equation}
where
dots denote higher order terms in ${\bf x}$. Importantly,  the
field $\d_i \d_j \eta_* ({\bf x})$ has blue power spectrum, unlike $\eta_*
({\bf x})$ and $\d_i \eta_* ({\bf x})$, so the major effect of the
infrared modes is accounted for by considering the two terms of the
gradient expansion written explicitly in (\ref{Eq/Pg12/1:cc_jetpl}).
Furthermore, we assume in what follows that $|\mathbf{v}|\ll 1$.
 The expansion in $|{\bf v}|$ is legitimate, since
the field ${\bf v} ({\bf x})$ has flat power spectrum
(cf. (\ref{Eqn/Pg11/1:cc_jetpl})), so the fluctuation of ${\bf v}$ is of
order $h^2 |\log \Lambda|$, where $\Lambda$ is the infrared cutoff, and it
is small for small $h$ and not too large  $|\log \Lambda|$. 

Keeping the two terms in (\ref{Eq/Pg12/1:cc_jetpl}) only, we have,
instead of (\ref{Eq/Pg3/2:cc_jetpl}),
\begin{equation}
(\delta \chi_2)^{\prime \prime}
- \d_i \d_i \; \delta \chi_2
- \frac{2}{[\eta_*(0) - {\bf vx} - \eta]^2} \delta \chi_2  = 0 \; .
\label{Eq/Pg12/2:cc_jetpl}
\end{equation}
We
observe that
the denominator in the expression for the background field
\begin{equation}
\chi_c = \frac{1}{h[\eta_* (0) - \eta - {\bf vx}]}
\label{Eq/Pg13/1:cc_jetpl}
\end{equation}
contains the combination
$\eta_* (0) - (\eta + {\bf vx})$.
We interpret this as the local time shift and Lorentz boost
of the original background (\ref{Eq/Pg3/1:cc_jetpl}).
Note that the field (\ref{Eq/Pg13/1:cc_jetpl}) is a solution to the field
equation
(\ref{Eq/Pg3/1A:cc_jetpl}) in our
approximation.
Our interpretation makes it clear that the solutions to
eq.~(\ref{Eq/Pg12/2:cc_jetpl}) can be obtained by time translation and
Lorentz boost of the original solution (\ref{Eq/Pg4/2:cc_jetpl}),
(\ref{Eq/Pg4/3:cc_jetpl}). In particular, instead of
(\ref{Eq/Pg5/2:cc_jetpl}) the phase field freezes out at
\begin{equation}
\delta \theta ({\bf x}) =
ih \int~\frac{d^3k}{4\pi^{3/2}\sqrt{k}(k+\mathbf{kv})}~
 \mbox{e}^{i{\bf kx} - i k\eta_*(\mathbf{x})}
\hat{A}_{\bf k}
+ h.c.
\label{Eq/Pg13/2:cc_jetpl}
\end{equation}

So far we discussed the dynamics of the phase perturbations $\delta
\theta $ at the conformal rolling stage which is governed solely be their
interaction with the background field (\ref{Eq/Pg13/1:cc_jetpl}) as well
as with the radial perturbations $\delta \chi _{1}$; the evolution of the
scale factor $a(\eta )$ is irrelevant.  After the end of conformal
rolling, the situation is reversed. Once the radial field $|\phi|$ has
relaxed to the minimum of the scalar potential, the phase $\theta$ is a
massless scalar field minimally coupled to gravity (this is true for any
Nambu--Goldstone field~\cite{voloshin}). Since we are talking about a yet
unknown pre-hot epoch, it is legitimate to ask what happens to the
perturbations of the phase right after the end of conformal rolling.
Barring fine tuning, there are two possibilities for the perturbations
$\delta \theta$:

(i) they are already superhorizon in the conventional sense at that time,
or

(ii) they are still subhorizon.

\subsection{Superhorizon phase perturbations}
\label{super-subs}

Let us consider the first sub-scenario~\cite{Libanov:2010nk, Libanov:2010ci}:
phase perturbations do not evolve after the end of the conformal rolling
stage, and the properties of the adiabatic perturbations are determined
entirely by the dynamics at conformal rolling (modulo possible
non-Gaussianity generated at the conversion epoch, see 
Sec.~\ref{Reprocess}).  This option is particularly
natural in the Galilean Genesis, but it is not contrived in the conformal
rolling scenario of Sec.~\ref{confroll} either.

In that case, there are no
potentially observable effects to the linear
order in $\mathbf{v}$. Indeed, using (\ref{Eq/Pg13/2:cc_jetpl}) one
finds for, e.g., two-point correlation function
\begin{eqnarray}
\langle \delta \theta (\mathbf{x})\delta \theta (\mathbf{x'})\rangle
&
\propto& \int~\frac{d^3 k}{k} \frac{1}{ (k + \vec{k}\vec{v})^2}
\mbox{e}^{i\vec{k}(\vec{x}- \vec{x}^\prime )
- ik{ (\eta_*(\vec{x}) - \eta_*(\vec{x}^\prime))}}
\nonumber
\\
&=& \int~\frac{d^3 q}{q} \frac{1}{ q^2}
\mbox{e}^{i\vec{q}(\vec{x}- \vec{x}^\prime )} ,
\label{Eqn/Pg14/1:cc_jetpl}
\end{eqnarray}
where  we have used (\ref{Eq/Pg12/1:cc_jetpl}) in the exponent and
 changed the integration variable from $\mathbf{k}$
to
\begin{equation}
\mathbf{q}=\mathbf{k}+k\mathbf{v}
\ ,\ \  q=|\mathbf{q}|=k+\mathbf{kv},
\label{Eq/Pg14/1:cc_jetpl}
\end{equation}
which is nothing but Lorentz boost. The result (\ref{Eqn/Pg14/1:cc_jetpl})
is precisely the two-point correlation function of the linear field
(\ref{Eq/Pg5/2:cc_jetpl}). The latter argument is straightforwardly
generalized to multiple correlators: for a given realization of the random
field $\eta _{*}(\mathbf{x})$, they are all expressed in terms of the
two-point correlation function (\ref{Eqn/Pg14/1:cc_jetpl}). The reason for 
the disappearance of the linear order effect is obviously Lorentz invariance.

The non-trivial effect of the large wavelength perturbations $\delta
\eta_* ({\bf x})$ on the perturbations of the phase, and hence on the
resulting adiabatic perturbations, occurs for the first time at the second
order in the gradient expansion, i.e., at the order $\partial _i \partial
_j \eta_*$~\cite{Libanov:2010nk}. Let us concentrate 
for the moment
on the effect of the
modes of $\delta \eta_*$ whose present wavelengths exceed the present
Hubble size. We are dealing with one realization of the random field
$\delta \eta_*$, hence at the second order of the gradient expansion,
$\partial _i \partial j \eta_*$ is merely a tensor, constant throughout
the visible Universe. In this long wavelength regime the perturbation of
the phase has the following form
\begin{equation}
\delta \theta ({\bf k}) =
\frac{ih\mbox{e}^{i{\bf kx} - i k\eta_*(\mathbf{x})}}{4\pi^{3/2}\sqrt{k}q}~
  \!\left(1\!-\!\frac{\pi }{2k}
\frac{k_{i}k_{j}}{k^{2}}\partial _{i}\partial _{j}\eta _{*}\! \right)\cdot
\hat{A}_{\bf k}
+ h.c.,
\label{Eq/Pg13/2a:cc_jetpl}
\end{equation}
where $q$ is given by (\ref{Eq/Pg14/1:cc_jetpl}). It results in the
power spectrum of the adiabatic perturbation which depends on
{\it directionality} of momentum
(see~\cite{Libanov:2010nk} for
details) 
\begin{align}
\mathcal{ P}_{\zeta }(\mathbf{k})=\mathcal{
P}^{0}(k)\!&\left(
1\! +\! c_1\! \cdot\! h\! \cdot \!\frac{H_0}{k} \cdot (n_k)_i
(n_k)_j w_{ij} \right.
\nonumber\\
& \left. - \!c_2 \cdot h^2 \cdot ({\bf  n_k u})^2 \phantom{\frac{A}{B}}\right).
\label{Eq/Pg15/1:cc_jetpl}
\end{align}
In the first non-trivial term, $w_{ij}$ is a traceless symmetric tensor of
a general form with unit normalization, $w_{ij} w_{ij} =1$, ${\bf n_k}$
is a unit vector, ${\bf n_k}= {\bf k}/k$, and $c_1$ is a constant of
order 1 whose actual value is undetermined because of the cosmic variance.
The last term is the result of the expansion of
$\delta \theta $ in $h$; the
corresponding term is not present in (\ref{Eq/Pg13/2a:cc_jetpl}). In this
term, ${\bf u}$ is a unit vector independent of $w_{ij}$, and the
positive parameter $c_2$ is logarithmically enhanced due to the infrared
effects,
$c_2 \propto  \log \frac{H_0}{\Lambda}$.
 This is the first place where the deep infrared modes show up.
Clearly, their effect is subdominant for small $h$.

We see that the large wavelength modes induce statistical anisotropy in
the adiabatic perturbations. The statistical anisotropy encoded in the
second term in (\ref{Eq/Pg15/1:cc_jetpl}) is similar to that commonly
discussed in inflationary context~\cite{aniso} (see 
Ref.~\cite{Shtanov} for earlier
analysis), and, indeed,
generated in some concrete inflationary models~\cite{soda}:  it does not
decay as momentum increases and has special tensorial form $({\bf n_k
u})^2$ with constant ${\bf u}$. On the other hand, the first non-trivial
term in (\ref{Eq/Pg15/1:cc_jetpl}) has the general tensorial structure and
decreases with momentum. The latter property is somewhat similar to the
situation that occurs in cosmological models with the anisotropic
expansion before inflation~\cite{Peloso}. 


Surprisingly, the statistical anisotropy in the form of the special
type quadrupole was found in the WMAP 5 and 7 years
data~\cite{GroeneboomEriksen,HansonLewis,GroeneboomEriksen2,RamazanovWMAP7}. It
was argued, however,
 that the anomaly may result from the detector beam
asymmetry not accounted for in the WMAP
analysis~\cite{HansonLewisCh}. In the final 9 years data release,
WMAP collaboration provided a set of maps deconvolved with the
instrument response function corresponding to the beam asymmetry
effect~\cite{WMAP9maps}. Deconvolved maps do not indicate the
statistical anisotropy allowing to constrain the coupling constant in
the first sub-scenario: $h^2 \log\frac{H_0}{\Lambda} <
1.2$~\cite{RamazanovWMAP9}. Later, the Planck data led to stronger
constraints on the statistical
anisotropy~\cite{Planck2013iso,KimKomatsu} and correspondingly to tighter
limits on the coupling constant, $h^2 \log\frac{H_0}{\Lambda}
< 0.30$~\cite{RamazanovPlanck}.


Another effect that emerges at order $h^{2}$ is 
the non-Gaussianity of the
perturbations  $\delta \theta$, and hence  the adiabatic
perturbations~\cite{ Libanov:2010ci,Libanov:2011bk}, 
over and beyond the non-Gaussianity that may be generated
at the time when the phase perturbations get reprocessed into the
adiabatic perturbations.
In the absence of the cubic self-interaction of the field $\theta$, the
intrinsic bispectrum vanishes, so we have to consider the 
trispectrum. It is fully calculated \cite{Libanov:2011bk},
the most striking feature being
the singularity in the limit where two momenta are
equal in absolute value and have opposite directions (folded limit, in
nomenclature of Ref.~\cite{Chen:2009bc}):
\begin{eqnarray}
\langle \zeta_{\mathbf{k}_{1}}  \zeta_{\mathbf{k}_{2}}
&\zeta_{\mathbf{k}_{3}}& \zeta_{\mathbf{k}_{4}} \rangle
= \mbox{const}  \cdot \delta \left(\sum
 \limits_{i=1}^{n}\mathbf{k}_{i}\right)\frac{1}{k_{12}k_1^{4}k_3^{4}}\nonumber
\\
&\times&  \left[1-3\left(
\frac{\mathbf{k_{12} k_1}}{k_{12} k_1} \right)^{2} \right]
\left[1-3\left(\frac{\mathbf{k_{12} k_3}}{k_{12} k_3} \right)^{2} \right]
\label{Ali/Pg16/1:cc_jetpl}\\
&&\mathbf{k_{12}} = \mathbf{k_{1}} + \mathbf{k_{2}} \to 0 \; , \nonumber
\end{eqnarray}
i.e., the trispectrum blows up as $k_{12}^{-1}$. This is in contrast to
trispectra obtained in single-field inflationary
models; indeed, there are general
arguments~\cite{Seery:2008ax} showing that in these models, the four-point
function is finite in the limit $k_{12} \to 0$. The singularity in the
four-point function (\ref{Ali/Pg16/1:cc_jetpl}) is due to the enhancement
of the radial perturbations $\delta |\phi |$ at low momenta.

Even though the results (\ref{Eq/Pg15/1:cc_jetpl}), (\ref{Ali/Pg16/1:cc_jetpl})
were first derived in the concrete model with negative quartic potential
(Sec.~\ref{confroll}), they are actually consequences of consistency
relations~\cite{Creminelli:2012qr} valid in the whole class of
conformal models. Furthermore, these consistency relations enable one
to calculate the one-loop non-Gaussianity in the folded limit
$k_{12} \to 0$. Interestingly the one-loop contribution to the trispectrum
is even more singular in the folded limit than the tree-level result
(\ref{Ali/Pg16/1:cc_jetpl}): one finds~\cite{Creminelli:2012qr}
\[
\langle \zeta_{\mathbf{k}_{1}}  \zeta_{\mathbf{k}_{2}}
\zeta_{\mathbf{k}_{3}} \zeta_{\mathbf{k}_{4}} \rangle^{(1-loop)}
= \mbox{const}  \cdot \delta \left(\sum
 \limits_{i=1}^{n}\mathbf{k}_{i}\right)\frac{1}{k_{12}^3k_1^{3}k_3^{3}} 
\log \frac{k_{12}}{\Lambda}\; 
\]
with suppressed coefficient (the suppression factor is $h^2$ in
the model of Sec.~\ref{confroll}). The $k_{12}^{-3}$-enhancement of the one-loop
trispectrum in the folded limit makes the non-Gaussianity 
even more promising from the
observational viewpoint.

\subsection{Subhorizon phase perturbations}

Let us consider another option: assume that there is a long enough period
of time after the end of conformal rolling, at which the phase
perturbations remain subhorizon in the conventional
sense~\cite{Libanov:2011hh}.  This option is fairly natural in the
conformal rolling model of Sec.~\ref{confroll} 
and more contrived in Galilean Genesis.

The behavior of $\delta \theta $ between the end of conformal rolling and
horizon exit depends strongly on the evolution of the scale factor at this
intermediate stage. In order that the flat power
spectrum (\ref{Eq/Pg5/3:cc_jetpl}) be not grossly modified at this epoch,
the scale factor should evolve in such a way that the dynamics of $\delta
\theta$ is effectively nearly Minkowskian. Although this requirement
sounds prohibitively restrictive, it is obeyed in 
the bouncing Universe, with matter
at the contracting stage having super-stiff equation of state, $p \gg
\rho$. It is worth noting in this regard that stiff equation of state is
preferred at the contracting stage for other reasons~\cite{smooth,ekpyro}
and is inherent, e.g., in a scalar field theory with negative exponential
potential, like in the ekpyrotic model~\cite{ekpyrosis}. It is
known~\cite{ekpyro-pert-old} that in models with super-stiff matter at
contracting stage, the resulting power spectrum of scalar perturbations is
almost the same as that of massless scalar field in Minkowski space,
${\cal P}(k) \propto k^2$. 
In tractable bouncing models like those of
Ref.~\cite{minus-exp}, the phase perturbations  evolve almost like 
in Minkowski space, exit
the horizon at the contracting stage, 
pass through the bounce unaffected
(cf. Ref.~\cite{Allen}), remain superhorizon early at the hot
expansion epoch and get reprocessed into adiabatic perturbations, as
discussed in Sec.~\ref{Reprocess}.

\begin{figure}[tb!]
\begin{center}
\includegraphics[width=0.4\textwidth,angle=0]{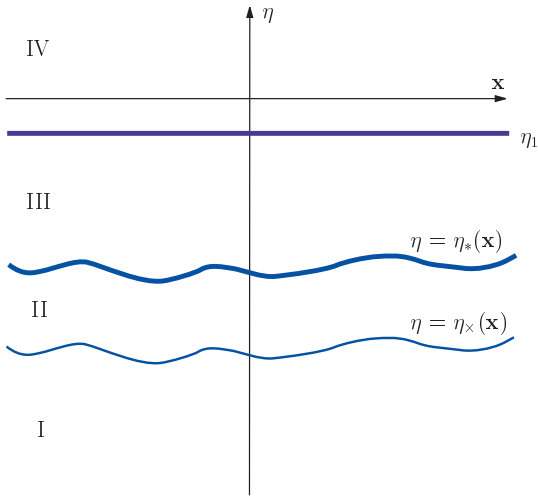}
\end{center}
\caption{{\rm Fig. 3}. Due to the perturbations of the radial field, the
evolution of phase perturbations proceeds in inhomogeneous background.
Perturbations $\delta \theta$ oscillate in time at early stage (region I),
freeze out at time $\eta=\eta_\times ({\bf x})$ and temporarily stay
constant (region II) until the end of conformal rolling that occurs at
$\eta=\eta_* ({\bf x})$. Then they evolve again, now in nearly Minkowskian
regime (region III), until the horizon exit time $\eta_1$.  Later on
(region IV), perturbations $\delta \theta$ are superhorizon and  stay
constant.
\label{Figure3}
 }
 \end{figure}
Let the effectively Minkowskian stage ends at some time $\eta _{1}$ (see
Fig.~\ref{Figure3}). The field $\delta
\theta ({\bf x}, \eta_*)$, determined by the dynamics at the conformal
rolling stage, serves as the initial condition for further Minkowskian
evolution from $\eta_*$ to $\eta_1$. We are interested in the properties
of the phase perturbations at $\eta = \eta_1$, as these properties are
inherited by the adiabatic perturbations.

To the leading order in $h$, we find nothing  new: the phase perturbations
at $\eta = \eta_1$ are Gaussian and have flat power spectrum. Subleading
orders in $h$ are more interesting. The effect of the perturbations
$\delta \eta_*(\mathbf{x})$ on the phase perturbations $\delta \theta$ is
twofold. First, the perturbations $\delta \eta_*$ modify the dynamics of
$\delta \theta$ at the conformal rolling stage. This property is the same as in
Sec.~\ref{super-subs}. 
Second, the
initial condition for the Minkwskian evolution
is now imposed at the non-trivial hypersurface $\eta =
\eta_* ({\bf x})$. This is illustrated in Fig.~\ref{Figure3}.
 The net result is that the perturbation $\delta \theta ({\bf x})$  at the
 time $\eta_1$ is a combination of two Gaussian random fields originating
from vacuum fluctuations of the phase $\theta$ and radial field $|\phi|$,
respectively.
This leads to potentially observable effects.

Let us consider the phase perturbation of given momentum
$\mathbf{k}$. 
At the end of conformal rolling it is given by (\ref{Eq/Pg13/2:cc_jetpl})
with $\partial \delta \theta/\partial \eta = 0$. This gives the initial
condition for the evolution at the intermediate stage, when the
perturbation is a linear combination of $\exp (i \mathbf{kx} \pm ik\eta)$.
So,
after the second freeze-out it is a linear combination of
waves coming from the direction  $\mathbf{n_{k}}=\mathbf{k}/k$ and from the
opposite direction and traveling distance $r=\eta _{1}-\eta _{*}$. This
leaves an imprint on $\delta \theta (\mathbf{k})$ of the random field
$\mathbf{v}$ existing at points $\mathbf{x} = \pm \mathbf{n_k} r$.
%
In particular,  
one finds the power spectrum with the non-trivial dependence on 
$\mathbf{n_{k}}$:
\[
{
\mathcal{ P}_{\delta \theta }(\mathbf{k})=\mathcal{ P}_{0}\left(1+{
\mathbf{n_{k}}}\cdot \left[{ \mathbf v(\mathbf{x} = 
+ \mathbf{n_{k}} r)}-{ \mathbf v(\mathbf{x}=-
\mathbf{n_{k}} r)} \right] \right) } \; .
\]
As a result, the
statistical anisotropy of the adiabatic perturbations has the form
\begin{equation}
{\cal P}_{\zeta} ({\bf k})= {\cal P}_\zeta^{(0)} (k) \left[1 +
Q(\mathbf{n_{k}})\right] \; ,
\label{Eq/Pg19/1:cc_jetpl}
\end{equation}
 where ${\cal P}_\zeta^{(0)}$ is independent of the directionality of momentum
 (nearly flat spectrum with small tilt)  and $ Q(\mathbf{n_{k}})$ is itself
a random field, which depends on the direction of ${\bf k}$ only. Unlike
the statistical anisotropy discussed in the inflationary
context~\cite{aniso,soda,Peloso} and also in Sec.~\ref{super-subs}, 
the function $ Q(\mathbf{n_{k}})$ contains all even
angular harmonics, starting from quadrupole.
We give here the expression
for $Q(\mathbf{n_{k}})$ which accounts for the quadrupole component only
(see Ref.~\cite{Libanov:2011hh} for all 
multipoles),
$Q(\mathbf{n_k}) =  {\cal Q} \cdot w_{ij} n_{\mathbf k}^i n_{\mathbf k}^j$
where $w_{ij}$ is a general symmetric traceless tensor normalized to
unity, $ w_{ij} w_{ij} = 1$, and the variance of the quadrupole component
(in the sense of an ensemble of universes) is
\begin{equation}
\langle {\cal Q}^2 \rangle = \frac{225 h^2}{32\pi^2}.
\label{Eq/Pg19/3:cc_jetpl}
\end{equation}
Of course, the precise values of the multipoles of  $ Q(\mathbf{n_{k}})$
in our patch of the Universe are undetermined because of the cosmic
variance. It is worth noting also that all multipoles are independent of
$k$ and, hence, unlike in the version of Sec.~\ref{super-subs}, 
there is no suppression of the
leading order effect on
cosmic microwave background (CMB) power spectrum at large
$l$.
 This property enables one to utilize the high statistics of the
Planck data
up to
 $l=1600$ and place a strong limit on the coupling constant in
 the sub-scenario with intermediate stage~\cite{RamazanovPlanck}: 
\begin{equation}
h^2<0.0011 \; .
\label{aug28-15-2}
\end{equation}
 The statistical anisotropy is
probably the most
promising signature of this sub-scenario.

The second effect is non-Gaussianity. While, as before, the
bispectrum vanishes,  the four-point correlation function has a peculiar
form
\begin{align}
\langle \zeta_{\bf k} &  \zeta _{\tilde{\bf k}} \zeta_{{\bf k}^\prime}
\zeta _{\tilde{\bf k}^\prime} \rangle
= \frac{  {\cal P}_\zeta^{(0)} \!(k)}{4\pi k^3} \frac{{\cal P}_\zeta^{(0)}
\!(k^\prime)}{4\pi k^{\prime \, 3}} \delta({\bf k}+\tilde{\bf k})
\delta({\bf k}^\prime+\tilde{\bf k}^\prime)
\nonumber \\
&
\times \left[ 1 + F_{NG} (\mathbf{n_k}\cdot \mathbf{n_{k^\prime}}) \right]
+ ({\bf k}
\leftrightarrow {\bf k}^\prime)  + (\tilde{\bf k} \leftrightarrow {\bf
k}^\prime) \; .
\label{Ali/Pg20/1:cc_jetpl}
\end{align}
The leading term in (\ref{Ali/Pg20/1:cc_jetpl}) (unity in square brackets)
is the Gaussian part, while the non-Gaussianity is encoded in
$F_{NG}=O(h^2)$. Note that the structure of the non-Gaussian part is
 fairly similar to that of the disconnected four-point function. Note also
that $F_{NG}$ depends on the angle between ${\bf k}$ and ${\bf k}^\prime$
only.
If the angle between ${\bf k}^\prime$
and ${\bf k}$ is small, i.e., $|\mathbf{n_k} -
 \mathbf{n_{k^\prime}}| \ll 1$,
the leading behavior of $F_{NG}$ is
\[
 F_{NG} = \frac{3h^2}{\pi^2} \log \frac{\mbox{const}}{|\mathbf{n_k} -
 \mathbf{n_{k^\prime}}|} \; ,
 \]
where constant in the argument of logarithm cannot be reliably
calculated because of the cosmic variance. The logarithmic behavior does
not hold for arbitrarily small $|\mathbf{n_k} -
 \mathbf{n_{k^\prime}}|$: the
function $F_{NG} (\mathbf{n_k} -
 \mathbf{n_{k^\prime}})$ flattens out most
likely at $|\mathbf{n_k} -
 \mathbf{n_{k^\prime}}| \sim [k(\eta_1 -
\eta_*)]^{-1/2}$, and certainly at $|\mathbf{n_k} -
 \mathbf{n_{k^\prime}}|
\sim [k(\eta_1 - \eta_*)]^{-1}$. So, the parameter $(\eta_1 - \eta_*)$ is
detectable in principle.

The third effect is negative scalar tilt
\begin{equation*}
n_s - 1 = - \frac{3h^2}{4\pi^2} \; .
\end{equation*}
However, this is not a particularly strong result, as small scalar
tilt in our scenario may also originate from explicit violation of
conformal invariance at the conformal rolling
stage~\cite{Osipov:2010ee} and/or not exactly Minkowskian evolution of
$\delta \theta$ at the intermediate stage. Moreover, 
to account for the whole scalar tilt detected by WMAP and Planck, one
needs $h \simeq 0.6$, in conflict with the constraint
(\ref{aug28-15-2}).

\section{Conclusions}

Flat or nearly flat power spectrum of the adiabatic perturbations may be a
consequence of  conformal symmetry rather than de Sitter symmetry. 
Models of this sort include conformally coupled
complex scalar field with negative quartic potential, Galilean Genesis and
decay of conformally invariant metastable vacuum. Properties of the
perturbations in these models are to large extent
dictated by conformal invariance and the
predictions are mostly model-independent, at least at the leading non-linear
level (modulo effects due to conversion of field fluctuations into
adiabatic perturbations). A peculiar property which has potentially
observable consequences is fluctuations along rolling direction. These
fluctuations have red power spectrum
and can be interpreted in terms of the local time shift.
Interplay between phase perturbations, responsible for density
perturbations in the end, and local time shift yields the non-trivial
correlation properties of the density perturbations such as statistical
anisotropy and the intrinsic non-Gaussianity of special
forms. The latter properties are potentially observable with CMB
data.

This work has been supported by Russian Science Foundation Grant No
14-12-01430.

\end{document}